\documentclass[aps,prl,twocolumn]{revtex4}  
\usepackage{graphicx}  
\usepackage{epsfig}
\usepackage{dcolumn}   
\usepackage{bm}        
\usepackage{amssymb}   
\usepackage{color}
\usepackage{cases}
\usepackage{subfigure}
\usepackage{float}
\begin{document}

\title{Zonal Flow Patterns: How Toroidal Coupling Induces Phase Jumps and Shear Layers}
\author{Z. B. Guo}
\altaffiliation[Corresponding author:]{guozhipku@gmail.com}
\author{P. H. Diamond}
\address{ University of California, San Diego, California 92093, USA}
\begin{abstract}
A new, frequency modulation mechanism for zonal flow pattern formation is presented. The model predicts the probability distribution function of the flow strength as well as the evolution of the characteristic spatial scale. Magnetic toroidicity-induced global phase dynamics is shown to determine the spatial structure of the flow. A key result is the observation that global phase patterning can lead to zonal flow formation in the absence of turbulence inhomogeneity.

\end{abstract}
\maketitle

How coherent structures emerge from turbulence is a central question of the physics of self organization. Examples of such structure formation phenomena include magnetic dynamo, vortex formation in shear layers and zonal flow formation in rotating or 2D fluids and magnetized plasmas\cite{Moffatt1978, Vallis2006, Diamond2005, Gurcan2015}. Zonal flow(ZF) dynamics is the topic of this paper, and is relevant to planetary atmosphere banding, jet stream formation and improved confinement in magnetized plasmas. The emergence of zonal flows in wave turbulence is frequently constrained by a type of adiabatic invariant, such as wave action. Adiabatic invariance follows from phase symmetry and scale separation\cite{Whitham2011}. Here the `phase'  refers to the global phase of the fluctuations. Specifically, an eikonal representation of each mode can be written as $\phi(\textbf{r},t)=|\phi(\textbf{r},t)|e^{iS(\textbf{r},t)}$ with the eikonal phase $S=\bar{S}+\tilde{s}$ composed of a slow, meso-scale piece $\bar{S}$--the global phase and a fast, micro-scale piece $\tilde{s}$. {{}For a homogeneous intensity($\nabla|\phi|=0$), the spatial variation of $\phi$ is also composed of a global piece($\nabla\bar{S}$) and a local piece($\nabla\tilde{s}$).} The evolution of the adiabatic invariant follows a global phase `trajectory' that extremizes the action of the underlying system\cite{Whitham2011}. Therefore, the space-time evolution of the global phase profile is central to structure formation and self-organization. Global phase dynamics has generally been overlooked in theories of zonal flow formation. In this work, we show that the geometry of the global phase which emerges from phase turbulence is intrinsic to broken symmetry in the turbulence, and hence is central to spatial structure formation. As a concrete example, important for both theory and application, we study global phase dynamics of ZF generation from drift wave turbulence in a toroidally confined plasma\cite{Fujisawa2008, Sarazin2010, Wang2006}. This study has as its principle outcomes the formulation of the theory of {\it frequency} modulational instability and global phase gradient shocks, along with the prediction of zonal flow formation in homogeneous turbulence($\nabla|\phi|^2=0$). A significant extension of the widely utilized Predator-Prey model of ZF dynamics is also proposed. 
     
ZFs play an essential role in regulating the level of turbulence and improving the energy confinement of magnetized plasmas. {{} Benefitting from progress in the state-of-art flux driven gyro-kinetic simulations, a more detailed spatial structure(i.e., quasi-`staircase' pattern) of the zonal flow in toroidal plasmas has been uncovered\cite{Dif2010, Dif2015}.} The ZF is driven by Reynolds' force. To produce a finite Reynolds' force, an inhomogeneous momentum flux is necessary. The most frequently invoked mechanism for ZF generation is {\it amplitude} modulation instability, where inhomogeneity of the momentum flux is induced by an initial modulation of the turbulence intensity by a `seed' ZF\cite{Diamond1998, Chen2000}. In this scenario, the ZF generated takes the general form, $\langle V\rangle(x,t)=\langle V\rangle_0(x)e^{\gamma_{ZF}t}$, where $\langle V\rangle_0$ is the seed ZF and $\gamma_{ZF}$ is the growth rate of the ZF. As is shown, the generated ZF has a memory of the structure of the seed, i.e., the spatial structure of the ZF is not entirely a derived quantity. By examining global phase dynamics, we show that the curvature of the global phase profile induces a contribution to the Reynolds' force, even when the turbulence intensity is homogeneous. ZF drive by global phase curvature is analogous to frequency modulation\cite{Roder1931}, for which the phase patterning induces a space-dependent global frequency. We also show that toroidicity-induced spatial phase coupling of the drift waves induces the formation of a global phase gradient `shock', --i.e., a layer with strong phase curvature. The shock layer then drives a ZF shear layer without a turbulence intensity gradient. The spatial structure and distribution of the ZF are determined by the corresponding properties of the phase shocks. This new picture also uncovers another role of toroidicity in plasma transport. In contrast to the common idea that magnetic toroidicity is unfavorable for confinement since it leads to ballooning\cite{Connor1978}, we show that toroidicity can, in fact, trigger phase gradient shocks and thus ZF formation. Incorporating the global phase evolution, we arrive at a {{} reaction-diffusion} system of coupled equations for ZF, turbulence intensity and global phase. The global phase is directly related to the cross phase of the momentum flux, so by inspecting how the cross phase evolves\cite{Yan2008}, one can follow the dynamics of the global phase, and vice versa.    
     
 We consider ZF evolution within a simple fluid picture
 \begin{equation}\label{ZF}
\frac{\partial}{\partial t}\langle V\rangle=-\frac{\partial}{\partial x}\langle v_xv_y\rangle-\gamma_d\langle V\rangle,
\end{equation}
where $\langle V\rangle$ is the poloidally averaged zonal flow velocity, $\langle v_xv_y\rangle$ is the Reynolds' stress($x$--radial direction, $y$- poloidal direction) and $\gamma_d$ is the ZF friction coefficient. $v=-\nabla \phi\times\hat{z}$ is the $E\times B$ drift velocity with $\phi$ the velocity stream function and proportional to the electrostatic potential. For simplicity, we take the toroidal mode number $n$ as fixed. After Fourier transformation in the poloidal direction, each poloidal mode can be written as $\phi_m=|\phi_m|e^{iS_m+im\theta}$ with $S_m=S_m(x,t)$ the eikonal phase of mode $m$. {{{}In a toroidally confined plasmas, the amplitude of each poloidal harmonic mode peaks at or near its associated rational surface, and is coupled with its neighbors via the toroidicity of the magnetic field. Thus, a quasi-periodic `chain'(i.e., quasi-lattice) is formed, with each $m$ corresponding to the radial position of a particular resonant surface(Fig.\ref{phaselattice}). A collective global oscillation can emerge due to couplings of the local harmonics\cite{Pikovsky2001}.} To explore the global phase dynamics in this lattice, the global phase function($\bar{S}$) is obtained by taking the continuum limit of the phase lattice(Fig.\ref{phaselattice}), so one has $S_m(x,t)=\bar{S}+\tilde{s}$. $\tilde{s}$ is the local phase, associated with each drift wave and $\partial_x\tilde{s}=k_x$ is the local radial wavenumber of the drift wave. Using the eikonal representation, $\phi_m$ can be written as $\phi_m=|\phi_m|e^{i\bar{S}+i\tilde{s}+im\theta}$.
\begin{figure}[htbp] 
\centering\includegraphics[width=3.5in]{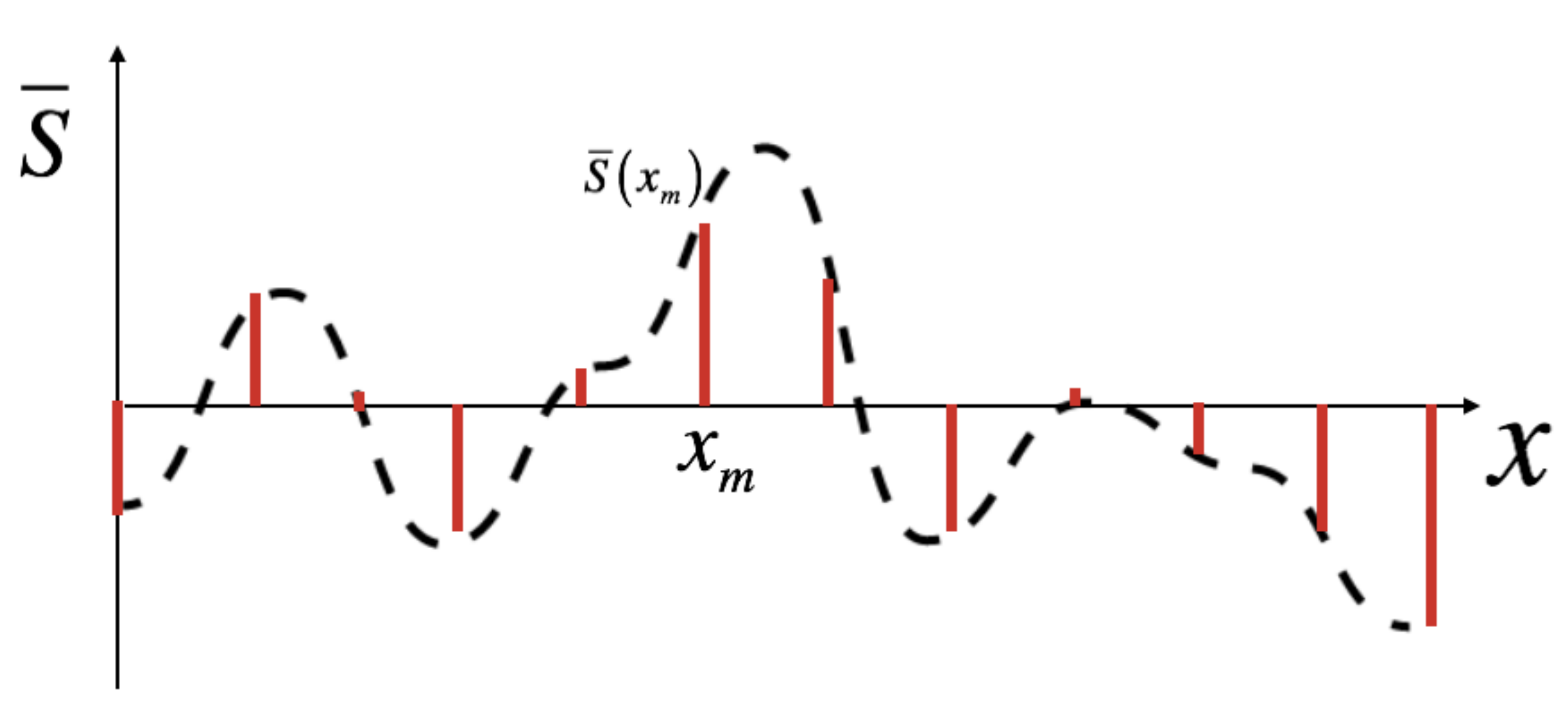} 
\caption{Red: phase lattice; dashed black: continuous limit of the phase lattice.}\label{phaselattice} 
\end{figure} 
The Reynolds' stress { {} at the resonance surface $x_m$} then follows as
\begin{equation}\label{Reynolds stress}
\langle v_xv_y\rangle=2\sum_{m'}{k_y'k_x'I(m')}+2\sum_{m'}{k_y'I(m')\frac{\partial}{\partial x}\bar S}
\end{equation}
where $I(m')\equiv |\phi_{m'}|^2/2$ is the intensity of the turbulence. $k_y$ is the poloidal wavenumber and is set by fast, small scales. In the continuum limit, $k_y$ can be understood as a continuous function of the radial position. $\partial_x \bar{S}$ can then be moved out of the summation, { {} since $\partial_x\bar{S}\simeq\partial_x\bar{S}|_{x=x_m}$}. For the 1st term to contribute, inhomogeneity of the turbulence intensity spectrum is required. In amplitude modulational stability, it is the seed ZF shear that modulates the turbulence intensity, inducing long range coherence of the turbulence, and hence inhomogeneity of the Reynolds' stress. { {} Note that since, after a reflection $m\rightarrow -m$, $k_y$ and $\bar{S}$ flip sign simultaneously}, the 2nd term in Eqn. (\ref{Reynolds stress}) is nonzero. So, we see that the global phase gradient can induce a finite cross correlation between $v_x$ and $v_y$ and hence a finite Reynolds' force if the global phase curvature is nonzero. {\it Note that this is the case even if the turbulence is homogeneous.} In other words, global phase curvature induces a {\it frequency modulation} mechanism, which is fundamentally different from the familiar amplitude modulation. One should note that Eqn. (\ref{Reynolds stress}) gives a general result for how the global phase pattern influences turbulent momentum transport. Using the spiky distribution and quasi-translation invariance(i.e., $k_x\simeq k_x'$) approximations at rational surfaces $x_m$, we need only consider contributions from the locally resonant mode $m$, i.e., $\sum_{m'}{...}\simeq \sum_{m'}{\delta_{mm'}...}$. Thus, Eqn. (\ref{ZF}) takes the form
\begin{equation}\label{ZF1}
\frac{\partial}{\partial t}\langle V\rangle\simeq 2k_{y}k_x\frac{\partial}{\partial x}I +2k_y\frac{\partial}{\partial x}I \frac{\partial}{\partial x}\bar S+2k_yI\frac{\partial^2}{\partial x^2}\bar{S}-\gamma_d\langle V\rangle,
\end{equation}
Note: the summation of the first three terms on the RHS is the total Reynolds' force and can be written in a conservative form, $\partial_x\left(2k_yk_xI+2k_yI\partial_x\bar{S}\right)$. The 1st term is the ZF acceleration driven by inhomogeneity of the turbulence intensity, which is the most familiar and frequently involved mechanism. The 2nd term is due to the combined effects of turbulence intensity inhomogeneity and the global phase gradient. The 3rd term is ZF acceleration by global phase curvature. This contributes even when the turbulence intensity is homogeneous, i.e., the global phase curvature itself can still induce a finite Reynolds' force and drive a ZF from zero. This new ZF drive mechanism is the most significant discovery of this paper. 

Focusing on this new mechanism, we consider ZF evolution when we assume the turbulence intensity to be homogeneous. The space-time structure of the turbulence intensity and its relation to global phase patterning are addressed later.   
ZF evolution driven by the global phase curvature follows as
\begin{equation}\label{ZF2}
\frac{\partial}{\partial t}\langle V\rangle= {2k_y}I\frac{\partial^2}{\partial x^2}\bar{S}-\gamma_d\langle V\rangle.
\end{equation}
To understand the mechanism  of the formation of the global phase curvature, one needs describe global phase evolution. A general way to obtain the global phase equation is by the eikonal equation
\begin{equation}\label{phase}
\frac{\partial}{\partial t}S=-\omega-\bold{k}\cdot\tilde{v},
\end{equation}
where $\omega=\omega_k+2\hat{\omega}_{De}+k_y\langle V\rangle$ is the total linear frequency, including its eigenfrequency($\omega_k$), magnetic drift frequency($2\hat{\omega}_{De}$) and the Doppler shift by the ZF. $\bold{k}\cdot\tilde{v}$ is the stochastic Doppler shift by the underlying turbulence. With $\bold{k}=\nabla \tilde{s}$, $\bold{k}\cdot\tilde{v}$ can be rewritten as $\bold{k}\cdot\tilde{v}=\nabla\cdot\Gamma_s$, where $\Gamma_s\equiv \bold{v}\tilde{s}$ is the turbulent phase flux. The magnetic drift frequency $2\hat{\omega}_{De}$ is a linear operator, and { {} $2\hat{\omega}_{De}\phi_m=\bold{v_d}\cdot\nabla\phi_m=V_D\left[k_y({\phi}_{m+1}+{\phi}_{m-1})-ik_x({\phi}_{m+1}-{\phi}_{m-1})\right]$ with $\bold{v_d}\cdot\nabla=V_D(k_y\text{cos}\theta+k_x\text{sin}\theta)$ the magnetic drift frequency and $V_D\equiv c_s\rho_s/R$\cite{Kadomtsev1970}}. 
In the continuum limit, and employing the strong coupling approximation($|k_x\Delta|\ll1$), one has $\phi_{m\pm1}\simeq [1\pm i\Delta\partial_x{S}-\frac{1}{2}(\Delta\partial_x{S})^2+...]\phi_m$ with $\Delta=1/(nq')$ the distance between rational surfaces at fixed $n$($q'$--gradient of the safety factor). The eigenvalue of $2\hat{\omega}_{De}$ follows as
\begin{equation}\label{De}
2\hat{\omega}_{De}\phi_m\simeq\left[2k_yV_D-k_yV_D\Delta^2\left(\frac{\partial {S}}{\partial x}\right)^2+{2k_xV_D\Delta\frac{\partial}{\partial x}{S}}\right]\phi_m. 
\end{equation} 
Eliminating the local, fast variation (i.e., $\partial_t\tilde{s}\simeq -\omega_k-2k_yV_D$) in Eqn. (\ref{phase}), the global phase evolution follows as
\begin{equation}\label{phase2}
\frac{\partial}{\partial t}\bar{S}\simeq-k_y\langle V\rangle-{2k_xV_D\Delta\frac{\partial}{\partial x}\bar{S}}+k_yV_D\Delta^2\left(\frac{\partial \bar{S}}{\partial x}\right)^2+D_s\frac{\partial^2}{\partial x^2}\bar{S},
\end{equation}
where, for closure, $\langle\Gamma_s\rangle$ is approximated by a Fickian flux with diffusion coefficient $D_s$,
\begin{equation}\label{Gamma}
\langle\Gamma_s\rangle=-D_s\frac{\partial}{\partial x}\bar S.
\end{equation}
Here $D_s\propto l_c^2\delta \omega$ with $l_c$ correlation length of the turbulence and $\delta \omega$ the turbulence decorrelation rate\cite{Xi2014}. 
The dynamics of the global phase is a consequence of four processes: frequency detuning by ZF(the 1st term on the RHS of Eqn. (\ref{phase2})), linear propagation(the 2nd term), quadratic self-coupling(the 3rd term), and turbulent diffusion(the 4th term). The frequency detuning strengthens the inhomogeneity of the phase profile. The linear propagation term can induce wave like propagation of the global phase profile, and the propagating velocity is $2k_xV_D\Delta$. In fact, by moving to a frame with radial velocity of $2k_xV_D\Delta$, the 2nd term in Eqn. (\ref{phase2}) can always be eliminated. The self-coupling term tends to induce nonlinear patterns in the phase profile and hence strengthens the phase curvature\cite{KPZ}. The turbulent diffusion term tends to flatten the phase profile, so driving it to saturation. Since the global phase always has a degree of gauge freedom, an equivalent quantity but one more symptomatic of turbulent mixing, is the global phase gradient. After applying a spatial derivative to both sides of Eqn. (\ref{phase2}), the evolution of the global phase gradient follows as   
\begin{eqnarray}\label{phase gradient}
\frac{\partial}{\partial t}\bar{S}'&=&-k_y\langle V\rangle'-{2k_xV_D\Delta\frac{\partial}{\partial x}\bar{S}'}+2k_xV_D\Delta^2\bar{S}'\frac{\partial}{\partial x}\bar{S}'\nonumber\\
&&+D_s\frac{\partial^2}{\partial x^2}\bar{S}',\label{phase gradient}
\end{eqnarray}
where $\bar{S}'\equiv {\partial_x}\bar{S}$. The 1st term on the RHS reflects the feedback effect of ZF shear on global phase gradient profile. Note that Eqn. (\ref{phase gradient}) is an inhomogeneous Burgers equation, and its most obvious property is the existence of a shock solution induced by the convective nonlinearity, $\bar{S}'\partial_x\bar{S}'$(Fig. \ref{phaseshock}).
\begin{figure}[htbp] 
\centering\includegraphics[width=3.5in]{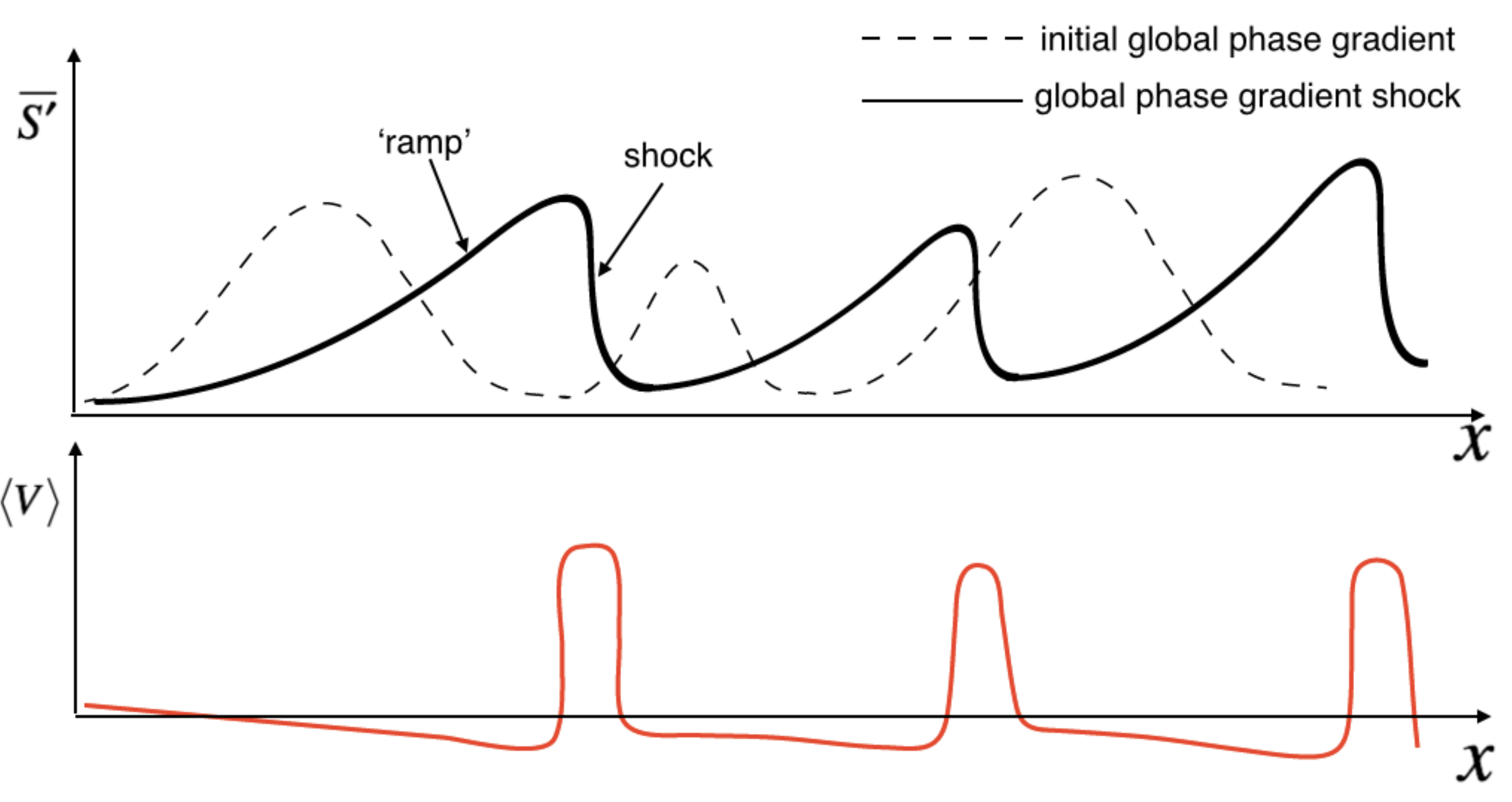} 
\caption{Top figure: global phase gradient shock induced by magnetic toroidicity; bottom figure: staircase like ZF bands induced by the shocks.}\label{phaseshock} 
\end{figure} 

In the initial stage, since the amplitude of the ZF is weak, we first ignore the feedback effect. In the shock layer region, the gradient of $\bar{S}'$(i.e., phase curvature) is large, so the ZF is strongly driven. In other words, the phase shock layer corresponds to a nascent ZF layer, and the width of the shock layer corresponds to the seed for the ZF shear length, $L_{ZF}$. The width of the shock layer is determined by the balance of the `overturning' effect, measured by the overturning time $|\delta \bar{S}'|/L_{ZF}$($\delta\bar{S}'$ is the jump in $\bar{S}'$ over the shock layer), and the diffusive effect (measured by the diffusion time $D_s/L_{ZF}^2$), i.e., 
\begin{equation}\label{shock}
2k_y V_D \Delta^2|\delta \bar{S}'|/L_{ZF}\simeq D_s/L_{ZF}^2,
\end{equation}     
where the jump of $\bar{S}'$ over the shock layer $\delta \bar{S}'$ is negative. We then obtain the scaling of the width of the shock layer
\begin{equation}
L_{ZF}\simeq \frac{D_s}{2k_y V_D \Delta^2|\delta\bar{S}'|}
\end{equation}
{{}$L_{ZF}$ can be further estimated by using the approximations: $D_s\simeq \rho_sc_s\rho_s/a$ for gyro-Bohm diffusion and $|\delta\bar{S}'|\simeq 1/\Delta$. Then, it follows as $L_{ZF}\simeq \frac{q'}{q}R\rho_s\simeq \frac{R}{a}\rho_s=\frac{R}{a}\frac{l_{meso}}{a}l_{meso}$ ($l_{meso}^2\equiv a\rho_s$ i.e., typical mesoscale), which is within the range observed in gyro-kinetic simulations\cite{Dif2010}. One can also see that the ZF shear($\langle V\rangle'$) deduced from Fig. (\ref{phaseshock}) exhibits a strong localized dipole structure, as was observed in \cite{Dif2015}.} 

The probability distribution function(PDF) of $\delta\bar{S}'$ depends on the `force' (i.e., the noise source). The noise originates from mode-mode beating processes\cite{MNR}. Incorporating a noisy forcing term $F(x,t)$, Eqn. (\ref{phase gradient}) becomes  
\begin{equation}\label{phase gradient2}
\frac{\partial}{\partial t}\bar{S}'= 2k_yV_D\Delta^2\bar{S}'\frac{\partial}{\partial X}\bar{S}'+D_s\frac{\partial^2}{\partial X^2}\bar{S}'+F(X,t),
\end{equation}
where $X\equiv x-2k_xV_D\Delta t$. The phase gradient-Burgers turbulence is an ensemble of `ramps'(phase gradient difference $\delta \bar{S}'>0$) and shocks($\delta\bar{S}'<0$)(Fig.\ref{phaseshock}). In the ramp region, the profile of $\bar{S}'$ is smooth, so that the phase curvature is small there and the ZF is not driven effectively. Thus, the ramps correspond to regions of fast transport. The shock regions are the site of ZF drive, and so correspond to local transport barriers. The alternating sequence of ramps and shocks resembles the staircase structure discovered in recent years, with the ramp corresponding to `step' and shock corresponding to the `jump' between steps(Fig.\ref{phaseshock}). 

By understanding the phase pattern, the spatial distribution of turbulent transport can be extracted. Assuming $F(x,t)$ to be noise which is white in time, $\langle F(x,t)F(x',t')\rangle\propto(x-x')^\xi\delta(t-t')$ with $\xi$ an index reflecting spatial inhomogeneity of the noise, we see that the PDF of the ramps follows an exponential scaling, $P(\delta\bar{S}'>0)\sim e^{-\delta\bar{S}^3/\delta\bar{S}_c^3}$\cite{Gurarie1996}, with $\delta S_c$ the characteristic value of the ramps. It is well known that the PDF of shocks has a power-law tail, as a consequence of the intermittency of the shock structures\cite{Chekhlov1995}. For homogeneous noise(i.e., $\xi=0$), one has $P(\delta\bar{S}'<0)\sim |\delta\bar{S}'|^{-4}$\cite{Chekhlov1995}. With Eqn. (\ref{shock}), one thus finds scaling of the PDF of the ZF width(or the width of the shock layer) to be
\begin{equation}\label{scaling}
P(L_{ZF})\sim L_{ZF}^{4}.
\end{equation}        
This power law scaling indicates the phase curvature driven ZF tends to concentrate at large scales. For inhomogeneous noise(i.e., $\xi\neq 0$ the external force is scale dependent), the index in Eqn. (\ref{scaling}) will be smaller than $4$, so, the shock layers tend to be sharper. The reason is that the external force will couple to the `inertial' range of the global phase gradient turbulence, so that it will inhibit formation of large shocks. The ZF generation and distribution is due to a roughening of the global phase profile\cite{Pikovsky2001}. The more roughening that occurs, the more curved the global phase profile will be, and so the ZF is more effectively driven at smaller scale. 

As the amplitude of the ZF develops to a certain value, one must consider its feedback on phase gradient evolution. An immediate observation is that the ZF shear tends to detune the phase gradient growth(i.e., via the 1st term in Eqn. (\ref{phase gradient})). This tends to enlarge the phase difference between neighbors in phase lattice, and hence enhances the roughness of the phase profile. In other words, the ZF shear has a positive feedback effect on the phase evolution. This feedback effect is most prominent near the `shoulder' of the phase gradient shock, where the ZF shearing rate is the strongest, and the overturning due to the nonlinear convection term is relatively weak. Thus, the ZF shear can be written as $\langle V\rangle'\simeq \left(-\partial_t\bar{S}'+D_s\partial_X^2\bar{S}'\right)/k_y$ and substituting into Eqn. (\ref{ZF2}) yields
\begin{equation}
\frac{\partial^2}{\partial t^2}\bar{S}'-\left(D_s\frac{\partial^2}{\partial X^2}-\gamma_d\right)\frac{\partial}{\partial t}\bar{S}'=\left(D_s\gamma_d-2k_y^2I\right)\frac{\partial^2}{\partial X^2}\bar{S}'.
\end{equation}  
After a Fourier transformation($\partial_t\rightarrow\gamma_\mathcal{K}$, $\partial_x\rightarrow i\mathcal{K}$), one has
\begin{equation}
\gamma_\mathcal{K}=\frac{\sqrt{(D_s\mathcal{K}^2-\gamma_d)^2+8k_y^2I\mathcal{K}^2}-(D_s\mathcal{K}^2+\gamma_d)}{2}.     
\end{equation}     
Existence of a positive growth rate requires
\begin{equation}
2k_y^2I>D_s\gamma_d,
\end{equation}     
i.e., distortion effect by ZF shear (measured by $2k_y^2I \mathcal{K}^2$) should exceed flattening effects by diffusion($D_s\mathcal{K}^2$) and damping by ZF friction($\gamma_d$). 

Due to the conservation of energy between ZF and turbulence, the appearance of ZF structures inevitably changes the spatial structure of the turbulence intensity, so that the initial assumption of homogeneity of $I$ ultimately fails. One must then consider the dynamical evolution of the turbulence intensity. The general form of the turbulence intensity evolution equation is
\begin{equation}\label{turbulence}
\frac{\partial}{\partial t}I=\gamma_l I+2k_yI\bar{S}'\langle V\rangle'+\frac{\partial}{\partial x}\left(D_TI\frac{\partial}{\partial x}I\right)-\gamma_{nl}I^2,
\end{equation}   
the 1st term is the linear driving term. The 2nd term comes from energy conservation between ZF and the turbulence. The 3rd term accounts for turbulence spreading\cite{Gurcan2004} with $D_TI$ the nonlinear turbulent intensity diffusion coefficient. The last term is a local turbulence dissipation/cascade term with $\gamma_{nl}$ the nonlinear dissipation coefficient. The specific forms of $D_T$ \& $\gamma_{nl}$ depend on the detailed properties of the underlying turbulence. As the details of these coefficients are the subject of this paper, we take $D_T$\&$\gamma_{nl}$ as given parameters. A positive linear growth is equivalent to the existence of a finite free energy flux(here the thermal energy flux), i.e., $\gamma_l\propto \langle v_x\tilde{P}\rangle$ with $\tilde{P}$ the thermal energy(pressure) fluctuation. Since $\langle v_x\tilde{P}\rangle \propto \text{cos}\theta_c$($\theta_c$ the cross phase between $v_x$ and $\tilde{P}$), the evolution of the turbulence intensity is necessarily coupled to the cross phase dynamics. Depending on the strength of ZF shearing, $\theta_c$ falls into two different states. For weak ZF shearing, $\theta_c$ is in a phase locked state, so that the turbulence is continuously pumped and the ZF is effectively driven by the phase curvature, as the phase curvature induced Reynolds' force is also proportional to the turbulence intensity. As ZF shear increases, the ZF will extract energy from the drift wave turbulence, and also upshift the cross phase, thus reducing the turbulence pumping rate\cite{Guo2015}. If the ZF shear exceeds a certain threshold, $\theta_c$ will enter and remain in a phase slip state, so the turbulence intensity will oscillate periodically. Then both the ZF and the global phase profile will exhibit quasi-periodic oscillation. Thus, due to the coupling to cross phase evolution, the dynamics of the ZF-global phase-turbulence intensity system is considerably enriched. Eqns. (\ref{ZF1}), (\ref{phase gradient})  and (\ref{turbulence}) constitute the {{} reaction-diffusion} system, where global phase steepening, flattening, ZF generation, and turbulence spreading are incorporated within a unified framework.           
     
In summary, we show that any minimal mean field theory of drift wave-ZF turbulence must include three players(not two): the ZF field, the turbulence intensity, and the global phase. By a new frequency modulation mechanism, the global phase curvature drives a ZF shear layer, {\it even when the turbulence is homogeneous}. Observe this mechanism explains the formation of robust ZF structure in core plasmas--in particular in regimes of modest mean $E\times B$ shear and weak turbulence inhomogeneity. A numerical test of this ZF formation mechanism is most viable. In a scenario with homogeneous turbulence and zero initial ZF field, by extracting the profile informations of global phase and ZFs, the dynamical relation(Eqns. (\ref{ZF2}) and (\ref{phase2})) between the global phase and the ZFs can be extracted directly. { {} In the current theory, the magnetic toroidicity induced {\it linear} mode coupling plays an essential role in inducing the global phase-gradient shock layer. It appears as a magnetic drift frequency($2\hat{\omega}_{de}$) in the eikonal equation, Eqn. (\ref{phase}). The turbulent diffusion effect is included, so that a steady shock layer can form. An important further study is whether {\it nonlinear} mode coupling can induce global phase pattern and then drive flow? This question is equivalent to studying whether the nonlinear frequency shift\cite{Tsytovich2012}, which can be added to the eikonal equation, can coherently couple the phases at different positions. The answer to this question depends on the detailed form of the nonlinear interaction and is definitely an important topic to pursue in the future.}              

\acknowledgments
We acknowledge fruitful interactions with participants in the 2015 Festival de Th\'{e}orie, Aix-en- Provence. We thank G. Tynan for stimulating discussions concerning experimental evidence for the role of the phase. This work was supported by the Department of Energy under Award Number DE-FG02-04ER54738 and CMTFO Award No. DE-SC0008378. Z. B. Guo was also supported by the Ministry of Science, ICT and Future Planning of the Republic of Korea under the Korean ITER project contract.

\end{document}